\documentstyle[prl,aps,multicol,epsfig]{revtex}

\input epsf.tex

\newcommand{\be}{\begin{equation}}
\newcommand{\ee}{\end{equation}}
\newcommand{\ba}{\begin{eqnarray}}
\newcommand{\ea}{\end{eqnarray}}
\newcommand{\ban}{\begin{eqnarray*}}
\newcommand{\ean}{\end{eqnarray*}}

\newcommand{\one}{\leavevmode\hbox{\small1\normalsize\kern-.33em1}}

\begin{document}

\title{Quantum cloning with an optical fiber amplifier}
\author{Sylvain Fasel, Nicolas Gisin, Gr\'egoire Ribordy, Valerio Scarani, Hugo Zbinden}
\address{
Group of Applied Physics, University of Geneva, 20, rue de
l'Ecole-de-M\'edecine, CH-1211 Geneva 4, Switzerland}
%\date{18 December 2001}
\maketitle

\begin{abstract}
It has been shown theoretically that a light amplifier working on
the physical principle of stimulated emission should achieve
optimal quantum cloning of the polarization state of light. We
demonstrate close-to-optimal universal quantum cloning of
polarization in a standard fiber amplifier for telecom
wavelengths. For cloning $1\rightarrow 2$ we find a fidelity of
0.82, the optimal value being $\frac{5}{6}=0.83$.
\end{abstract}

\begin{multicols}{2}

Classical information can be copied at will. Not so for the
information content of a quantum state: one cannot devise a
process that takes $N$ copies of an arbitrary quantum state as an
input, and produces $M>N$ copies of the same quantum state
deterministically. This is the content of the {\em no-cloning
theorem} of quantum mechanics \cite{noclon}, which is at the heart
of quantum information theory (in particular, it guarantees the
security of quantum cryptography). To go beyond this no-go
theorem, one can weaken the requirements, and ask that the $M$
copies are not identical to the input state, but as close as
possible to it \cite{hilbuz,others}. The physical device that
performs this operation is called {\em quantum cloning machine}. A
device that copies equally well all the possible input states is
called universal quantum cloning machines (UQCM).

In the recent years, communication through optical fibers has
become widespread, and everybody knows that a light signal can be
amplified. But light can (should) be described
quantum-mechanically, therefore the standard amplification devices
used in telecom cannot beat the no-cloning theorem.

It is not difficult to understand why some noise will always be
produced by the amplifier: the amplification of light is achieved
through {\em stimulated emission}, and it is well-known that in
this case {\em spontaneous emission} will always be present as
well. But it has been noticed recently \cite{simon} that the
amplification based on stimulated emission leads to {\em optimal
cloning}. To see this, describe the amplifier as an ensemble of
atoms initially in the excited state, that can emit photons
polarized either along any direction with equal cross-section. The
atoms are irradiated with a photon of the suitable energy,
polarized along a direction $V$. At the exit of the amplifier, we
select the cases in which one and only one additional photon has
been emitted, and analyze the output in the $(H,V)$ basis. If $p$
is the probability that the additional photon is polarized along
$H$ (spontaneous emission), then the probability that the
additional photon is polarized along $V$ is $2p$ because of
stimulated emission. Now, if we pick one photon of the output at
random, the probability of this photon to be in the same state as
the input photon (namely $V$) is called {\em fidelity} of the
cloner. In the case that we are considering, the fidelity is
$\frac{5}{6}$, which is indeed the optimal fidelity for a
$1\rightarrow 2$ universal cloning machine \cite{hilbuz}. This
easy reasoning has been extended to any amplification process
$N\rightarrow M$ in Ref. \cite{simon}; we re-derive the main
results below.

According to this theoretical prediction, an amplifier whose gain
is independent of the polarization is a UQCM for the polarization
states of photons. Amplification through parametric
down-conversion has been considered \cite{simon,dema,oxf,other}.
In this Letter, we demonstrate an amplification in an Er-doped
fiber that is very close to optimal cloning.

We begin by reviewing some theoretical elements on cloning and
amplification, while stressing the links with our experiment. The
setup itself and the results are described in detail in the second
half of the Letter.

\begin{center}
\begin{figure}
\epsfxsize=7cm \epsfbox{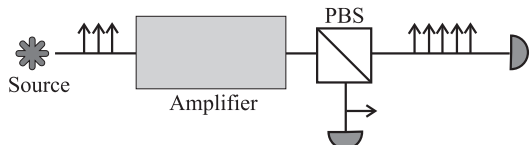} \caption{A cloning
experiment: a state of $N=3$ photons is amplified to a state of
$M=6$ photons. Spontaneous emission may create photons in the
wrong polarization mode. PBS: polarizing beam-splitter.}
\label{expideal}
\end{figure}
\end{center}

{\em Cloning of polarization states.} An experiment to demonstrate
universal cloning of polarization states consists of three blocks
(fig. \ref{expideal}): the preparation, the amplification
(cloning) and the analysis. The source prepares $N$ photons in the
same polarization mode, say $V$. The photons are sent into the
amplifier, supposed to be non-birefringent to ensure that any
input polarization is amplified in the same way (universal
cloning). Suppose that at the output of the amplifier one selects
the events in which {\em exactly} $M>N$ photons have been
produced. According to the no-cloning theorem, it is impossible
that all $M$ photons are deterministically in the state $V$: some
of the photons at the output have been produced in the orthogonal
mode $H$ because of spontaneous emission, and will consequently be
reflected at the polarizing beam splitter (PBS). Thus the process
$N\rightarrow M$ is characterized by the probabilities $p_M(k|N)$
that the $M$ output photons are distributed as: $N+k$ photons in
the mode $V$, and $M-N-k$ photons in the orthogonal mode $H$, with
$0\leq k\leq M-N$: \ba p_M(k|N)&\equiv&\mbox{Prob}[(N+k)_V,
(M-N-k)_H|N_V,0_H]\,. \ea We normalize these probabilities so that
$\sum_k p_M(k|N)=P(M|N)$, the probability of the process
$N\rightarrow M$. The fidelity of the process is defined as the
fraction of photons that are found in the same mode as the input:
\ba {\cal{F}}_{N\rightarrow
M}&=&\frac{N+\bar{k}_{NM}}{M}\label{deffid} \ea with
$\bar{k}_{NM}\,=\,\sum_{k=0}^{M-N} k\,\frac{p_M(k|N)}{P(M|N)}$. If
the amplification process is based on stimulated emission with no
absorption, then all the $p_M(k|N)$ are proportional to the
probability of the spontaneous emission $p_M(0|N)$ through the
binomial factor \cite{note1} \ba \frac{p_M(k|N)}{p_M(0|N)}
&=&\frac{(N+k)!}{N!\,k!}\,,\;1\leq k\leq M-N\,. \ea Inserting
these probabilities into (\ref{deffid}), one recovers exactly the
optimal fidelity for a cloning $N\rightarrow M$ \cite{others}: \ba
{\cal{F}}_{N\rightarrow M}^{opt}&=&\frac{MN+M+N}{M(N+2)}\,.
\label{fidopt}\ea Note that this result is independent of $P(M|N)$
or $p_M(0|N)$: these quantities are in general difficult to
calculate, which means that one doesn't know how frequent the
process $N\rightarrow M$ is (see \cite{simon} for estimates in
some limiting cases). Nevertheless, each process that takes place
would show the optimal fidelity if it could be isolated from the
other processes.

{\em Photon statistics.} The two-dimensional quantum degree of
freedom (qubit) that we want to clone is the polarization of
photons. More precisely, one qubit corresponds to one photon per
mode. Our source does not produce a Fock state of $N$ photons, but
a continuous light signal, with weak power $P_{in}$. Its spectral
density is centered at the frequency $\nu$ and has a width width
$\Delta\nu$. In this context, the concept of photon is introduced
as the {\em energy quantum}: writing $P_{in}=\mu_{in}
h\nu/\tau_c$, with $\tau_c\propto\Delta\nu^{-1}$ the coherence
time, we see that the input power corresponds to an average of
$\mu_{in}$ photons per spatio-temporal mode, that is per coherence
time. Our source produces states of $n$ photons, this number being
statistically distributed with a distribution $p(n)$, with average
$\sum_n np(n)=\mu_{in}$. In principle, one could then use a fast
photon detector to count the number of photons per time-modes, but
this is not possible in practice for the coherence time used in
our experiment. However, a measurement of the intensity is a
direct way of measuring the mean values of the photon statistics.

The input light is polarized along a direction that we label $V$.
After the amplification stage, the PBS allows the measurement of
the intensities in each polarization mode, that is the mean
numbers of photons $\mu_V$ and $\mu_H$. The fidelity is defined as
above: the fraction of photons that is found in the same
polarization mode as the input light, that is \cite{note2} \ba
\bar{{\cal{F}}} \,=\,\frac{\mu_V}{\mu_{out}}\,,\;&\mbox{ with }&
\;\mu_{out}=\mu_{V}+\mu_{H}\,. \label{fidmoy}\ea In other words,
we are performing an experiment on light amplification in the weak
intensity regime. Can one extract information about the underlying
quantum cloning processes from such a measurement?

A great insight is gained by describing our experiment with a
semi-classical theory of light amplification. Since we measure
only mean intensities, we can simply take eq. (14) in the seminal
paper by Shimoda {et al.} \cite{japan}, and write it in our
notations for each of the modes $V$ and $H$: \ba
\mu_V\,=\,G\mu_{in}+\frac{1}{Q}(G-1)\;&\;,\;&\;
\mu_H\,=\,\frac{1}{Q}(G-1)\,. \label{musjapan}\ea The two
parameters $G$ and $Q$ are not independent, but are determined by
the microscopic details of the process. $G$ is the gain due to
stimulated emission \cite{gainnote}; $Q$ can be used as a figure
of merit for the UQCM. In fact: $Q=1$ means no absorption, in
which case we know (see above and \cite{simon}) that all
underlying processes have the optimal fidelity. When $Q=0$, the
absorption compensates exactly the emission; in this case, we have
also $G=1$. This means that the gains and losses in the amplifier
compensate each other, and all the additional intensity
$\mu_{out}-\mu_{in}$ comes from spontaneous emission. This is
obviously the worst possible cloning machine \cite{qneg}.

The formulas (\ref{musjapan}) relate the gain $G$ to $Q$,
$\mu_{in}$ and $\mu_{out}$ as $G=(Q\mu_{out}+2)/(Q\mu_{in}+2)$.
Inserting this into the fidelity (\ref{fidmoy}), we obtain \ba
\bar{{\cal{F}}}_{\mu_{in} \rightarrow \mu_{out}}&=&
\frac{Q\mu_{out}\,\mu_{in}
+\mu_{out}+\mu_{in}}{Q\mu_{out}\mu_{in}+2\mu_{out}}
\label{fidq}\ea Note that for $Q=1$ the r.h.s. is formally the
same as the optimal fidelity ${\cal{F}}_{\mu_{in}\rightarrow
\mu_{out}}^{opt}$ (\ref{fidopt}), but here $\mu_{in}$ and
$\mu_{out}$ need not be integers. For instance, if $Q=1$,
$G=\frac{4}{3}$ and $\mu_{in}=1$, we have $\mu_{out}=2$ and
$\bar{{\cal{F}}}_{\mu_{in} \rightarrow
\mu_{out}}=\frac{5}{6}={\cal{F}}_{1\rightarrow 2}^{opt}$. In
conclusion: in the absence of absorption, the mean fidelity is the
optimal fidelity for the mean numbers of photons. This somewhat
astonishing result is a new manifestation of the deep link between
the classical and the quantum description of light that has been
stressed in a recent historical review of laser physics
\cite{lamb}.

\begin{center}
\begin{figure}
\epsfxsize=8cm \epsfbox{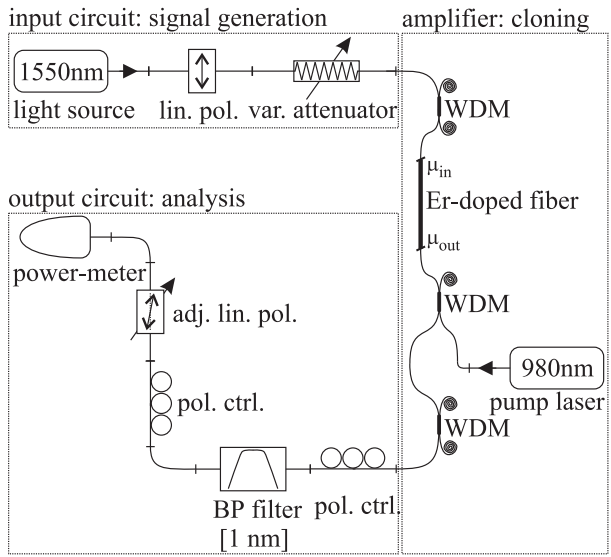} \caption{Scheme of the
setup. See text for details.} \label{expreal}
\end{figure}
\end{center}

{\em The setup.} We proceed to the detailed description of the
experimental setup. In the scheme (fig. \ref{expreal}), one
recognizes the realization of each of the three blocks:
preparation, amplification and analysis.

To prepare the polarized photons, we use a source of unpolarized
light \cite{note3} followed by a linear polarizer that achieves an
extinction ratio of about 21dB between the two orthogonal
polarizations. An adjustable attenuator is then used in order to
tune the power. This attenuator is also useful to prevent the
light coming from the amplifier to be back-reflected into the
circuit, which would create a hardly controllable ghost signal.

The spectrum of the source is wide; a band-pass tunable filter can
be used to reduce the spectral width to the desired value
$\Delta\nu$ around the working wavelength $\frac{c}{\nu}\approx
1550$nm. This tunable filter is actually placed after the
amplifier so that both the signal and the amplified light are
filtered through it. This is not a nuisance since the light at
different wavelengths does not disturb the process of
amplification (this is because we inject a very low power compared
to the saturation level of the amplifier). The filter sets the
width of the optical mode $\Delta\nu$, thus defining the power
corresponding to one photon per mode.

The second block of the setup is the amplifier (the cloning
machine), which consists of a few tens of centimeters of pumped
Erbium-doped fiber (EDF). We note that a commercial amplifier
(consisting of meters of EDF) would not be suitable for our
experiment, since it is optimized to achieve a gain much higher
than the ones we want. The pump is a 980nm laser with output power
120mW, thus making the fiber an inverted medium capable of
amplifying a signal around 1550nm. The pumping is done backward
with respect to the signal in order to limit residual pump at the
output. Since the pump and the signal have different wavelengths,
the separation of the signal from the pump is done by wavelength
division multiplexers (WDM) at both ends of the EDF. The WDM
between the source and the EDF is used to avoid pump light to
disturb or destroy the source apparatus. At the other end of the
EDF we put two WDMs, the second one acting as a filter for the
light which is back-reflected from the first one.

The third block is the analyzer. It consists of an adjustable
linear polarizer, together with a polarization controller and a
single power-meter. With the polarization controller, one can
align the setup so that on of the axes of the adjustable polarizer
corresponds to the polarization of the input signal, while the
orthogonal mode is the "noise".

{\em Measurement protocol and results.} Before starting the
experiment, one must optimize the working wavelength, align the
analyzer, and determine the losses in the circuit in order to
calibrate the measurement of $\mu_{in}$ and $\mu_{out}$.

The working wavelength is chosen with the tunable pass-band
filter, with a width in wavelength of about 1nm. It is determined
experimentally at 1555nm by searching, within the range of the
filter, the best emission-over-absorption ratio, i.e. the
wavelength where the absorption is minimized but the gain is not
zero. The alignment of the polarization controller in the analyzer
is performed by generating a signal at the source but leaving the
pump off.

The losses in the circuit must be determined precisly because the
relevant experimental quantities to demonstrate cloning are the
power at the entry of the EDF, giving $\mu_{in}$, and the power
corresponding to each polarization mode at the exit of the EDF,
giving $\mu_{V}$ and $\mu_H$. The polarization-dependent loss of
the whole circuit is due mainly to the filter; we measure it using
the fluorescence of the pumped EDF without signal --- by the way,
this light is found to be totally depolarized, meaning that all
the polarizations will be cloned equally well as desired. The
losses in the analyzing block, including the two WDMs, are
measured using a tunable laser to avoid measuring losses due to
the reduction of the spectral width. We note that the fidelity
$\bar{\cal{F}}$ calculated using (\ref{fidmoy}) does not depend on
the losses nor on the error on the losses, because these are
multiplicative factors that cancel out in the division. Thus the
estimation of $\bar{\cal{F}}$ can be made with high precision.

The power at the entry of the EDF is calibrated using the signal
from the source, with the adjustable attenuator set to a reference
value, and of course without pumping the EDF. At the analysis
power-meter, we measure the power corresponding to the
spectrum-window defined by the filter, from which we must deduce
the losses in the output circuit and inside the fiber. For this
calibration the absorption inside the EDF itself must be precisely
determined. We found an attenuation of 0.25dB. With this
procedure, we know the value of $\mu_{in}$ corresponding to each
position of the adjustable attenuator.

For the experiment, the pump is turned on. The input power is
scanned using the adjustable attenuator. For each mean power
$\mu_{in}$ in the input, the mean power corresponding to $\mu_{V}$
(resp. $\mu_H$) is determined by reading the value at the
power-meter when the polarizer is aligned along the input state
(resp. its orthogonal state), and deducing only the losses of the
analyzing block.

The length and the doping of the EDF have been chosen in order to
achieve the desired gain at the working wavelength. The
measurements presented here where made on a commercial EDF (INO
Er103), 37cm long. With these values, we have a mean number of
photons in the output $\mu_{out}\approx 1.94$ for a mean number of
photons in the input $\mu_{in}\approx 1$.

\begin{center}
\begin{figure}
\epsfxsize=10cm \epsfbox{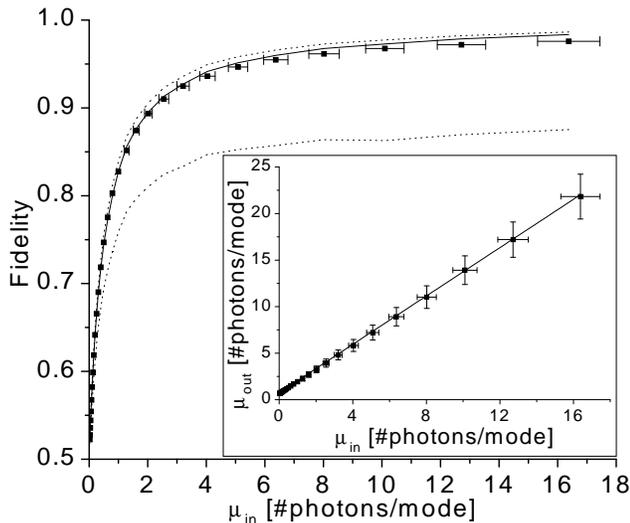} \caption{Inset:
$\mu_{out}$ as a function of $\mu_{in}$; the linear fit allows to
extract $G$ and $Q$ defined in (\ref{musjapan}). Main figure:
fidelity (\ref{fidmoy}) as a function of $\mu_{in}$. Solid line:
$Q=0.8$, best fit with eq. (\ref{fidq}). Dotted lines: upper:
$Q=1$ (optimal cloning); lower: $Q=0$ (no cloning).}
\label{results}
\end{figure}
\end{center}

The experimental results are shown in fig. \ref{results}. In the
inset, we show that a linear relation holds between $\mu_{out}$
and $\mu_{in}$ for all the input powers, in agreement with
formulas (\ref{musjapan}). From our data, we extract the
values of the two parameters $G$ and $Q$. We find $G=1.3$ and
$Q=0.8$.

In the main part of the figure, we show the data for the fidelity
calculated from (\ref{fidmoy}), as a function of the mean number
of photons in the input. The solid line correspond to
eq. (\ref{fidq}) with $Q=0.8$. The dotted lines correspond
respectively to the optimal cloner ($Q=1$, upper line) and the
worst cloner ($Q=0$, lower line). The experimental curve is
clearly close to the optimal cloner, which confirms that $Q$, the
parameter describing the absorption in the amplifier, is indeed a
good figure of merit.

In conclusion, we have demonstrated close-to-optimal quantum
cloning of the polarization state of light using a standard fiber
amplifier working on the physical principle of stimulated
emission. Since the amplifier is not birefringent, it acts as a
universal cloning machine. On the side of application: An
universal cloner is the optimal device for an eavesdropper to
attack the six-state protocol of quantum cryptography
\cite{sixstate}, while a better strategy can be chosen to attack
the four-state protocol \cite{niu}. The results of this Letter
show that the physical realization of this device is not a very
hard step; it will however be much harder for Eve to store the
photons and wait for Alice and Bob to reveal the bases
\cite{felix}. On the fundamental side, we like to conclude by
stressing again the discussion about eq. (\ref{fidq}). Quantum
cloning could have been noticed and measured in the early days of
laser physics; but it was not, because the notion of {\em
information} was not yet central in science and consequently the
quantum community was not aware of the fundamental role of the
concept of (im)possible copying.

\end{multicols}


\begin{thebibliography}{10}

\bibitem{noclon} W.K. Wootters, W.H. Zurek, Nature {\bf 299}, 802
(1982); P.W. Milonni, M.L. Hardies, Phys. Lett. A {\bf 92}, 321
(1982)
\bibitem{hilbuz} V. Bu\v{z}ek, M. Hillery, Phys. Rev. A {\bf 54},
1844 (1996)
\bibitem{others} N. Gisin, S. Massar, Phys. Rev.
Lett. {\bf 79}, 2153 (1997); D. Bruss et al., Phys. Rev. Lett.
{\bf 81}, 2598 (1998)
\bibitem{simon} C. Simon et al., Phys. Rev. Lett.
{\bf 84}, 2993 (2000); J. Kempe et al., Phys. Rev. A {\bf 62},
032302 (2000)
\bibitem{dema} F. De Martini et al., Opt. Comm. {\bf 179}, 581
(2000)
\bibitem{oxf} A. Lamas-Linares et al., Science {\bf 296}, 712 (2002)
\bibitem{other} Y.-F. Huang et al., Phys. Rev. A {\bf 64}, 012315 (2002); H.K.
Cummins et al. Phys. Rev. Lett. {\bf 88}, 187901 (2002). These two
experiments are not based on amplification, but on transferring
part of the information onto a different degree of freedom
initially in a blank state.
\bibitem{note1} This can easily be seen by modelling the amplifier as a
chain of atoms that start all in the excited state, and having the
initial photon distribution evolving according to the following
rule: If an atom is irradiated by $n$ V-photons and $m$ H-photons,
the probabilities of emitting one additional photon in either mode
are linked by $p(+1_{V}|n,m)/p(+1_{H}|n,m)=(n+1)/(m+1)$ --- and
there is of course a probability $1-p(+1|n,m)$ that no additional
photons are emitted by the atom.
\bibitem{note2}  In terms of the individual processes, the average fidelity reads
 \ban \bar{{\cal{F}}}&=&\frac{\sum_n p(n)\,\sum_M
P(M|n)\,M\,{\cal{F}}_{n\rightarrow M}}{\sum_n p(n)\,\sum_M
P(M|n)\,M}\,. \ean However, this expression is hard to estimate,
because we don't know the $P(M|n)$.
\bibitem{japan} K. Shimoda et al., J. Phys.
Soc. Japan {\bf 12}, 686 (1957). For the process that we consider,
$a=c$ holds in the notations of the paper.
\bibitem{gainnote} That is, the
ratio between [$\mu_{out}$-spontaneous emission] and $\mu_{in}$.
\bibitem{qneg} We don't even consider the case $Q<0$, which would
mean that the absorption overrates the emission, that is, that the
amplifier is actually an absorber!
\bibitem{lamb} W.E. Lamb et al.,
Rev. Mod. Phys. {\bf 71}, S263 (1999)
\bibitem{note3} As a source, we use the spontaneous emission of a
commercial Erbium-doped fiber amplifier.
\bibitem{sixstate} D. Bruss, Phys. Rev. Lett. {\bf 81}, 3018 (1998);
H. Bechmann-Pasquinucci, N. Gisin, Phys. Rev. A {\bf 59}, 4238
(1999)
\bibitem{niu} C.-S. Niu, R.B. Griffiths, Phys. Rev. A {\bf
60}, 2764 (1999); V. Scarani, N. Gisin, Phys. Rev. Lett. {\bf 87},
117901 (2001)
\bibitem{felix} S. F\'elix et al., J. Mod. Opt. {\bf 48}, 2009
(2001)


\end{thebibliography}
\end{document}